\documentclass[a4paper,11pt]{article}

\usepackage{mathtools}
\usepackage{amsthm}
\usepackage{amsmath}
\usepackage{amsfonts}
\usepackage{enumitem}
\usepackage{bm}
\usepackage{authblk}

\newcommand{\Rm}[1]{\mathop{\mathrm{#1}}\nolimits}

\newcommand{\FF}{\mathbb{F}}

\newcommand{\0}{\mathbf{0}}

\DeclareMathOperator{\wt}{wt}

\newtheorem{thm}{Theorem}
\newtheorem{lem}{Lemma}
\newtheorem{cor}{Corollary}

\title{On the existence of quaternary Hermitian LCD codes with Hermitian dual distance $1$}
\author{Keita Ishizuka\thanks{Corresponding author. Research Center for Pure and Applied Mathematics Graduate School of Information Sciences, Tohoku University, Sendai 980–8579, Japan. email: \texttt{keita.ishizuka.p5@dc.tohoku.ac.jp}}, Ken Saito\thanks{Research Center for Pure and Applied Mathematics Graduate School of Information Sciences, Tohoku University, Sendai 980–8579, Japan. email: \texttt{kensaito@ims.is.tohoku.ac.jp}}}
\date{}

\begin{document}
\maketitle

\begin{abstract}
    For $k \ge 2$ and a positive integer $d_0$, we show that if there exists no quaternary Hermitian linear complementary dual $[n,k,d]$ code with $d \ge d_0$ and Hermitian dual distance greater than or equal to $2$, then there exists no quaternary Hermitian linear complementary dual $[n,k,d]$ code with $d \ge d_0$ and Hermitian dual distance $1$.
    As a consequence, we generalize a result by Araya, Harada and Saito on the nonexistence of some quaternary Hermitian linear complementary dual codes.
    \\
    \textbf{Keywords.} Linear code, Linear complementary dual code, Hermitian linear complementary dual code.
    \\
    \textbf{2010 AMS Classification.} 94B05
\end{abstract}

\section{Introduction}
\label{sec:Introduction}
% LCD符号の定義
Let $\FF_q$ be the finite field of order $q$, where $q$ is a prime power.
Let $\FF_q^n$ be the vector space of all $n$-tuples over $\FF_q$ and let $x = (x_1, x_2, \ldots, x_n)$ be a vector of $\FF_q^n$.
The Hamming weight of $x$ is defined as $\wt(x) = \# \{ i \mid x_i \neq 0 \}$.
In this note, we are concerned only with linear codes.
Thus we omit the term ``linear'' and simply write ``code''.
A $k$-dimensional subspace $C$ of $\FF_q^n$ is said to be an $[n, k]$ code over $\FF_q$.
The parameters $n, k$ are said to be the length, dimension of $C$ respectively.
A vector of $C$ is said to be a codeword of $C$.
The codes over $\FF_2, \FF_3, \FF_4$ are said to be the binary, ternary, quaternary codes respectively.
Let $C$ be an $[n, k]$ code over $\FF_q$ and let $\0_n$ be the zero vector of $\FF_q^n$.
The minimum weight of $C$ is defined as $d(C) = \min \{ \wt(x) \mid x \in C,\,x \neq \0_n \}$.
If the minimum weight of $C$ equals to $d$, then $C$ is said to be an $[n, k, d]$ code over $\FF_q$.

% LCD codes
Let $x = (x_1, x_2, \ldots, x_n),\, y = (y_1, y_2, \ldots, y_n) \in \FF_q^n$.
For $\FF_q^n$, the Euclidean inner product is defined as $(x, y) = \sum_{i=1}^n x_i y_i$.
The Euclidean dual code $C^{\perp}$ of an $[n, k]$ code $C$ over $\FF_q$ is defined as $C^{\perp} = \{ x \in \FF_q^n \mid (x,y) = 0 \textrm{ for all } y \in C \}$.
Let $x = (x_1, x_2, \ldots, x_n),\, y = (y_1, y_2, \ldots, y_n) \in \FF_{q^2}^n$.
For $\FF_{q^2}^n$, the Hermitian inner product is defined as $(x, y)_h = \sum_{i=1}^n x_i \overline{y_i}$, where $\overline{y_i} = y_i^q$.
The Hermitian dual code $C^{\perp h}$ of an $[n, k]$ code $C$ over $\FF_{q^2}$ is defined as $C^{\perp h} = \{ x \in \FF_{q^2}^n \mid (x,y)_h = 0 \textrm{ for all } y \in C \}$.
A code $C$ is said to be a Euclidean (resp.\ Hermitian) linear complementary dual code, a Euclidean (resp.\ Hermitian) LCD code for short, if $C \cap C^{\perp} = \{ \0_n \}$ (resp.\ $C \cap C^{\perp h} = \{ \0_n \}$).
The Euclidean (resp.\ Hermitian) dual distance $d^{\perp}$ (resp.\ $d^{\perp h}$) of $C$ is the minimum weight of the Euclidean (resp.\ Hermitian) dual code.

Carlet, Mesnager, Tang, Qi and Pellikaan~\cite{carlet2018linearcodes} proved that any code over $\FF_{q^2}\,(q > 2)$ is equivalent to a quaternary Hermitian LCD code.
Lu, Li, Guo and Fu~\cite{lu2014maximal} proved that a quaternary Hermitian LCD code leads to a construction of a maximal-entanglement entanglement-assisted quantum error correcting code.
For $k \ge 2$ and a positive integer $d_0$, Araya, Harada and Saito proved that if there exists no binary (resp.\ ternary) Euclidean LCD $[n,k,d]$ code with $d \ge d_0$ and $d^\perp \ge 2$, then there exists no binary (resp.\ ternary) Euclidean LCD $[n,k,d]$ code with $d \ge d_0$ and $d^\perp=1$~\cite[Proposition 3.3]{araya2020characterization}.
This motivates us to study the existence of quaternary Hermitian LCD codes with $d^{\perp h}=1$.
In this note, it is shown that a result analogous to~\cite[Proposition 3.3]{araya2020characterization} holds for quaternary Hermitian LCD codes.

The remaining of this note is organized as follows.
In Section~\ref{sec:Preliminaries}, we recall preliminary results and prove a lemma needed later.
In Section~\ref{sec:TheExistenceOfQuaternaryHermitianLCDCodesWithDualDistance}, for $k \ge 2$ and a positive integer $d_0$, it is shown that if there exists no quaternary Hermitian LCD $[n,k,d]$ code with $d \ge d_0$ and $d^{\perp h} \ge 2$, then there exists no quaternary Hermitian LCD $[n,k,d]$ code with $d \ge d_0$ and $d^{\perp h}=1$.
Furthermore, we apply the result and obtain a generalized version of~\cite[Theorem 9]{araya2020quaternary}.

% TODO: sectionとともにlabelを自動生成するsnippet作る
\section{Preliminaries}
\label{sec:Preliminaries}
A generator matrix of a code $C$ is any matrix whose rows form a basis of $C$. 
Given a matrix $G$, we denote the transpose, conjugate of $G$ by $G^T, \overline{G}$ respectively. 
% DONE: Hermitianだけでいい？
The following characterization on Hermitian LCD codes is due to G\"{u}neri, B. \"{O}zkaya and P. Sol\'{e}~\cite{guneri2016quasi}.
\begin{lem}[{\cite[Proposition 3.5]{guneri2016quasi}}]
    \label{lem:LCDCharGenmat}
    Let $C$ be an $[n, k]$ code over $\FF_{q^2}$ and let $G$ be a generator matrix of $C$.
    Then $C$ is a Hermitian LCD code if and only if the $k \times k$ matrix $G\overline{G}^T$ is nonsingular.
\end{lem}
\begin{lem}
    \label{lem:dualDistReduced}
    Let $k \ge 2$.
    If there exists a quaternary Hermitian LCD $[n, k, d_0]$ code with $d^{\perp h} \ge 2$, then there exists a quaternary Hermitian LCD $[n+1, k, d]$ code with $d^{\perp h} \ge 2$ and $d \in \{ d_0, d_0+1 \}$.
\end{lem}
\begin{proof}
    Let $C_0$ be a quaternary Hermitian LCD $[n, k, d_0]$ code with $d^{\perp h} \ge 2$.
    From the proof of \cite[Proposition 4]{harada2020remark}, $C_0$ has a generator matrix of the form
    \begin{math}
        G_0 = 
        \begin{pmatrix}
            x \\
            G_1 \\
        \end{pmatrix}
    \end{math},
    satisfying that $( x, x )_h \neq 0$ and $x \overline{G_1}^T = \0_{k-1}$.
    The code $C_1$ with generator matrix $G_1$ is a quaternary Hermitian LCD $[n, k-1]$ code.
    \begin{enumerate}[label=$(\mathrm{\roman*})$]
        \item $k = 2$: $G_1$ is regarded as a codeword $c_1 \in C_1$ with $( c_1, c_1 )_h \neq 0$ since $C_1$ is a Hermitian LCD code.
        \item $k \ge 3$: From the proof of \cite[Proposition 4]{harada2020remark}, $C_1$ has a generator matrix of the form
            \begin{math}
                G_1' = 
                \begin{pmatrix}
                    x' \\
                    G_2' \\
                \end{pmatrix}
            \end{math},
            satisfying that $( x', x' )_h \neq 0$ and $x' \overline{G_2'}^T = \0_{k-2}$.
            The code $C_2'$ with generator matrix $G_2'$ is a quaternary Hermitian LCD $[n, k-2]$ code.
            Since both $G_1$ and $G'_1$ are generator matrices of $C_1$,
            \begin{align*}
                G'_0
                = \begin{pmatrix}
                    x \\
                    G'_1 \\
                \end{pmatrix}
                = \begin{pmatrix}
                    x \\
                    x' \\
                    G'_2 \\
                \end{pmatrix}
            \end{align*}
            is also a generator matrix of $C_0$.
            Since $x \in C_1^{\perp h}$, we have $( x, x' )_h = 0$ and $x G_2'^T = \0_{k-2}$.
    \end{enumerate}

    Consider the following $k \times (n+1)$ matrix $G$:
    \begin{align*}
        G = 
        \begin{cases}
            \begin{pmatrix}
                G_0 & h^T
            \end{pmatrix} & \Rm{if} k = 2, \\
            \begin{pmatrix}
                G_0' & h^T
            \end{pmatrix} & \Rm{if} k \ge 3, \\
        \end{cases}
    \end{align*}
    where
    \begin{align*}
        h =
        \begin{cases}
            (1, 1) & \Rm{if} k = 2, \\
            (1, 1, \0_{k-2}) & \Rm{if} k \ge 3. \\
        \end{cases}
    \end{align*}
    Let $C$ be the code with generator matrix $G$.

    \begin{enumerate}[label=$(\mathrm{\roman*})$]
        \item $k = 2$: it holds that
            \begin{equation*}
                \det G \overline{G}^T
                % &= \det
                % \begin{pmatrix}
                %     x & 1 \\
                %     G_1 & 1 \\
                % \end{pmatrix}
                % \begin{pmatrix}
                %     \overline{x}^T & \overline{G_1}^T \\
                %     1 & 1 \\
                % \end{pmatrix} \\
                = \det
                \begin{pmatrix}
                    x \overline{x}^T + 1 & x \overline{G_1}^T + 1 \\
                    G_1 \overline{x}^T + 1 & G_1 \overline{G_1}^T + 1 \\
                \end{pmatrix} \\
                % &= \det
                % \begin{pmatrix}
                %     0 & 1 \\
                %     1 & 0 \\
                % \end{pmatrix} \\
                = 1.
            \end{equation*}

          \item $k \ge 3$: it holds that
              \begin{equation*}
                  \det G \overline{G}^T
                  % &= \det
                  % \begin{pmatrix}
                  %     x & 1 \\
                  %     x' & 1 \\
                  %     G_2' & \0_{k-2} \\
                  % \end{pmatrix}
                  % \begin{pmatrix}
                  %     \overline{x}^T & \overline{x'}^T & \overline{G_2'}^T \\
                  %     1 & 1 & \0_{k_2}^T \\
                  % \end{pmatrix} \\
                  = \det
                  \begin{pmatrix}
                      x \overline{x}^T + 1 & x \overline{x'}^T + 1 & x \overline{G_2'}^T \\
                      x' \overline{x}^T + 1 & x' \overline{x'}^T + 1 & x' \overline{G_2'}^T \\
                      G_2' \overline{x}^T & G_2' \overline{x'}^T & G_2' \overline{G_2'}^T \\
                  \end{pmatrix} \\
                  % &= \det
                  % \begin{pmatrix}
                  %     0 & 1 & \0_{k-2} \\
                  %     1 & 0 & \0_{k-2} \\
                  %     \0_{k-2} & \0_{k-2} & G_2' \overline{G_2'}^T \\
                  % \end{pmatrix} \\
                  = \det G_2' \overline{G_2'}^T \\
                  \neq 0.
              \end{equation*}
        \end{enumerate}
        In any case, $G\overline{G}^T$ is nonsingular.
        By Lemma~\ref{lem:LCDCharGenmat}, $C$ is a Hermitian LCD code.
        Furthermore, it holds that for all nonzero codeword $c \in C$ there exists a nonzero codeword $c_0 \in C_0$ and $a \in \FF_4$ such that $c = (c_0, a)$.
        Therefore, it follows that $d(C) \in \{ d_0, d_0+1 \}$.
        It is clear that $d^{\perp h} \ge 2$.
        Hence $C$ is a quaternary Hermitian LCD $[n+1, k, d]$ code with $d^{\perp h} \ge 2$ and $d \in \{ d_0, d_0+1 \}$.
    \end{proof}

\section{The existence of quaternary Hermitian LCD codes with Hermitian dual distance $1$}
\label{sec:TheExistenceOfQuaternaryHermitianLCDCodesWithDualDistance}
    \begin{thm}
        \label{thm:dualDistOne}
        Let $k \ge 2$ and let $d_0$ be a positive integer.
        If there exists no quaternary Hermitian LCD $[n, k, d]$ code with $d \ge d_0$ and $d^{\perp h} \ge 2$,
        then there exists no quaternary Hermitian LCD $[n, k, d]$ code with $d \ge d_0$ and $d^{\perp h} = 1$.
    \end{thm}
    \begin{proof}
        Suppose that there exists a quaternary Hermitian LCD $[n, k, d]$ code $C$ with $d \ge d_0$ and $d^{\perp h} = 1$.
        % DONE: zero coordinatesは用例あり．W. C. Huffman and V. Pless, Fundamentals of Error-correcting codes, page 282.
        Let $l$ be the number of zero coordinates of $C$.
        Puncturing $C$ on the zero coordinates gives a quaternary Hermitian LCD $[n-l, k, d]$ code with $d^{\perp h} \ge 2$.
        Then a quaternary Hermitian LCD $[n, k, d]$ code with $d \ge d_0$ and $d^{\perp h} \ge 2$ is constructed by Lemma~\ref{lem:dualDistReduced}.
        This contradicts the assumption.
    \end{proof}

    For positive integers $n,k,\alpha$, define $r_4(n,k,\alpha) = 4^{k-1}n - \frac{4^k-1}{3}\alpha$.
    Araya, Harada and Saito~\cite{araya2020quaternary} gave the following result on the nonexistence of some quaternary Hermitian LCD codes with $d^{\perp h} \ge 2$.
    \begin{thm}[{\cite[Theorem 9]{araya2020quaternary}}]
        \label{thm:AHS}
        Let $k \ge 3$ and $4\alpha - 3n \ge 1$.
        \begin{enumerate}[label=$(\mathrm{\roman*})$]
            \item Suppose that $4r_4(n,k,\alpha) < k$.
                Then there exists no quaternary Hermitian LCD $[n, k, \alpha]$ code with $d^{\perp h} \ge 2$.
            \item Suppose that $4r_4(n,k,\alpha) \ge k$.
                If there exists no quaternary Hermitian LCD $[4r_4(n,k,\alpha), 3r_4(n,k,\alpha)]$ code with $d^{\perp h} \ge 2$, then there exists no quaternary Hermitian LCD $[n, k, \alpha]$ code with $d^{\perp h} \ge 2$.
        \end{enumerate}
    \end{thm}
    
    By Theorem~\ref{thm:dualDistOne}, we remove the condition on the Hermitian dual distance as follows.
    \begin{cor}
        \label{cor:r4}
        Let $k \ge 3$ and $4\alpha - 3n \ge 1$.
        \begin{enumerate}[label=$(\mathrm{\roman*})$]
            \item Suppose that $4r_4(n,k,\alpha) < k$.
                Then there exists no quaternary Hermitian LCD $[n, k, \alpha]$ code.
            \item Suppose that $4r_4(n,k,\alpha) \ge k$.
                If there exists no quaternary Hermitian LCD $[4r_4(n,k,\alpha), 3r_4(n,k,\alpha)]$ code with $d^{\perp h} \ge 2$, then there exists no quaternary Hermitian LCD $[n, k, \alpha]$ code.
        \end{enumerate}
    \end{cor}

    Araya, Harada and Saito determined the largest minimum weight among all quaternary Hermitian LCD $[n,3]$ codes~\cite[Theorem 12]{araya2020quaternary}.
    In the proof of~\cite[Theorem 12]{araya2020quaternary}, the nonexistence of some quaternary Hermitian LCD $[n,3]$ codes with $d^{\perp h} \ge 2$ was established by Theorem~\ref{thm:AHS}.
    On the other hand, the nonexistence of some quaternary Hermitian LCD $[n,3]$ codes with $d^{\perp h} =1$ was established by case-by-case analyses.
    By Corollary~\ref{cor:r4}, we only have to consider quaternary Hermitian LCD $[n,3]$ codes with $d^{\perp h} \ge 2$.
    This simplifies the proof.

    \section*{Acknowledgement}
    The authors are grateful to Professor Masaaki Harada for bringing this problem to their attention.


\begin{thebibliography}{99}
        % \bibitem{araya2019onthe} M. Araya and M. Harada, On the classification of linear complementary dual codes, Discrete Math.\ 342 (2019), 270--278.
        \bibitem{araya2020characterization} M. Araya, M. Harada and K. Saito, Characterization and classification of optimal LCD codes, Des.\ Codes Cryptogr. 89 (2021), 617--640.
        \bibitem{araya2020quaternary} M. Araya, M. Harada and K. Saito, Quaternary Hermitian linear complementary dual codes, IEEE Trans.\ Inform.\ Theory 66 (2020), 2751--2759.
        % \bibitem{bosma1997magma} W. Bosma, J. Cannon and C. Playoust, The Magma algebra system.\ I. The user language, J. Symbolic Comput.\ 24 (1997), 235--265.
        % \bibitem{carlet2016complementary} C. Carlet and S. Guilley, Complementary dual codes for counter-measures to side-channel attacks, Adv.\ Math.\ Commun.\ 10 (2016), 131--150.
        \bibitem{carlet2018linearcodes} C. Carlet, S. Mesnager, C. Tang, Y. Qi and R. Pellikaan, Linear codes over $\FF_q$ are equivalent to LCD codes for $q > 3$, IEEE Trans.\ Inform.\ Theory 64 (2018), 3010--3017.
        \bibitem{guneri2016quasi} C. G\"{u}neri, B. \"{O}zkaya and P. Sol\'{e}, Quasi-cyclic complementary dual codes, Finite Fields Appl.\ 42 (2016), 67--80.
        \bibitem{harada2020remark} M. Harada and K. Saito, Remark on subcodes of linear complementary dual codes, Inform.\ Process.\ Lett.\ 159 (2020), 105963, 3 pp. 
        % \bibitem{huffman2010fundamentals} W. C. Huffman and V. Pless, Fundamentals of Error-Correcting Codes, First Edition, Cambridge University Press (2010).
        \bibitem{lu2014maximal} L. Lu, R. Li, L. Guo and Q. Fu, Maximal entanglement entanglement-assisted quantum codes constructed from linear codes, Quantum Inf.\ Process.\ 13 (2015), 165--182.
        % \bibitem{massey1992linear} J. L. Massey, Linear codes with complementary duals, Discrete Math.\ 106/107 (1992), 337--342.
\end{thebibliography}
\end{document}